# Broadband Cyclic-Symmetric Magnet-less Circulators and Theoretical Bounds on their Bandwidth

Ahmed Kord, *Student Member, IEEE*, Dimitrios L. Sounas, *Senior Member, IEEE*, Zhicheng Xiao, *Student Member*, *IEEE*, and Andrea Alù, *Fellow*, *IEEE*

*Abstract*— In this paper, we explore theoretically and experimentally broadband spatiotemporally modulated (STM) magnet-less circulators realized by combining three-port non-reciprocal junctions with three identical bandpass filters. We develop a rigorous theory for the proposed circuit, which allows to optimize its design and to derive a global bound on the maximum possible bandwidth (BW). We verify our theory with simulations and measurements of a printed circuit board (PCB) prototype based on a differential wye junction and second-order Chebyshev bandpass filters, resulting in a measured fractional BW of 13.9% at a center frequency of 1 GHz.

## I. INTRODUCTION

RECIPROCITY is a general principle governing the fact that signal transmission between two points in space is identical in either direction regardless of the complexity of the intermediate channel. It applies to a broad range of materials, under the conditions that they are non-gyrotropic, passive, linear, and time-invariant [1]-[3]. Breaking any of these conditions allows the implementation of non-reciprocal components, such as circulators, isolators, and gyrators, which have numerous applications in modern communication systems [4]-[10]. Recently, magnetless implementations of such components based on linear periodically time-varying circuits [11]-[38] have received significant attention, as they were shown to overcome the weight, size, and cost challenges of magnetic devices [39]-[43], while satisfying the requirements on all other essential metrics. In particular, [28] and [29] presented radio-frequency (RF) circulators that can simultaneously achieve low transmission loss, excellent matching, large isolation, high power handling and linearity, low noise figure, and good harmonic response, all at a small form factor and low modulation frequency and amplitude. These circuits are also compatible with standard CMOS technologies, which permits further integration and cost reduction for large-scale production. Nevertheless, this remarkable performance was maintained over a relatively narrow fractional bandwidth (BW) of only a few percent. While this BW may be sufficient for a plethora of applications such as WiFi, RFID, and cellular communication, it is still highly desirable to increase it further for broadband scenarios, such as radar systems and future ultra-wideband radios.

The described STM circulators have been based on connecting bandpass or bandstop resonators in a wye or a delta topology, respectively, while modulating their natural oscillation frequencies with a particular phase pattern [28]. The instantaneous BW in either topologies is limited by the modulation parameters (particularly the modulation frequency), the order of the constituent resonators, and the loaded quality factor. It was shown in [28], [29] that first-order resonators, i.e., series or parallel *LC* tanks, with 10-20% modulation frequency and 50 Ohm termination can lead to a BW of 3~4% at best in practice. In order to improve this further, one solution is to increase the modulation frequency, yet this would increase power consumption, prohibit integration using thick-oxide CMOS technologies which can handle high power, and complicate scaling the center frequency to the mm-wave band. Another solution is to use transformers to downconvert the typical 50 Ohm impedance of the ports to a smaller value, thus decreasing the loaded quality factor and broadening the junction's resonance. However, the required modulation amplitude in this case may become unrealistic, since it has to increase when the loaded *Q* decreases in order to maintain large isolation (IX) and small insertion loss (IL) [28]. Alternatively, one could increase the order of the constituent resonators so that the junction supports more modes (two degenerate modes at each pole) which, consequently, can be designed to provide isolation over a larger BW. While this is indeed a viable option, the main challenge is that it increases the circuit complexity.

A simpler approach to broaden the bandwidth of magnetless STM circulators was introduced in [32] based on

This paper is an expanded version of an abstract presented at the 2018 International Microwave Symposium, Philadelphia, PA, USA, 10-15 June 2018. The authors are with the Department of Electrical and Computer Engineering, University of Texas at Austin, Austin, TX 78712, U.S.A. A. A. is also with the Photonics Initiative, Advanced Science Research Center, with the Physics Program, Graduate Center, and with the Department of Electrical Engineering, City College of New York, City University of New York, New York, NY 10031, U.S.A. (corresponding author: A. A., +1.212.413.3260; e-mail: aalu@gc.cuny.edu). This work was supported by the IEEE Microwave Theory and Techniques Society Graduate Fellowship, the Qualcomm Innovation Fellowship, the Air Force Office of Scientific Research, the Defense Advanced Research Projects Agency, Silicon Audio, the Simons Foundation, and the National Science Foundation. A. A. is currently the Chief Technology Officer of Silicon Audio RF Circulator. The terms of this arrangement have been reviewed and approved by The University of Texas at Austin and the City University of New York in accordance with its policy on objectivity in research.

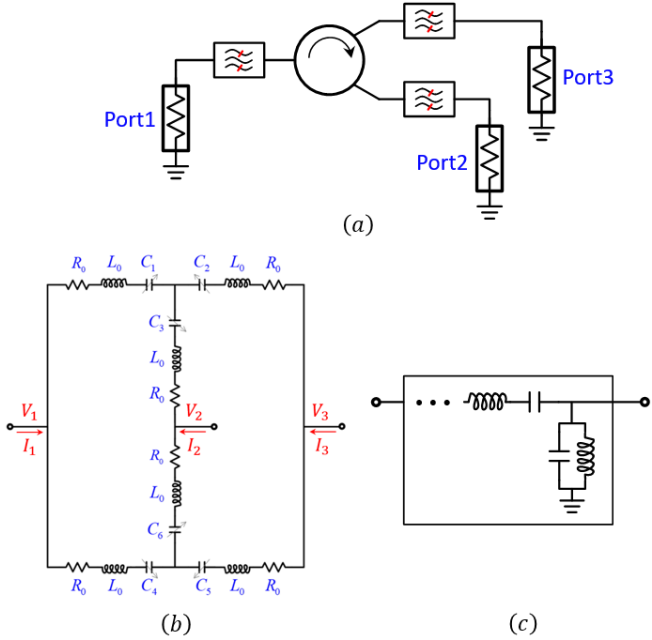

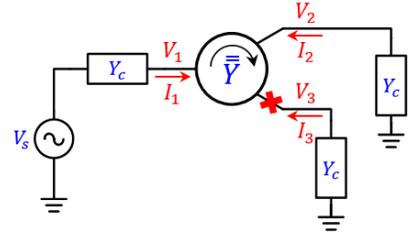

Fig. 1. (a) Proposed broadband magnetless circulator. (b) STM differential current-mode junction [29]. DC and modulation networks are not shown for simplicity. (c) Ladder bandpass filter.

combining narrowband non-reciprocal junctions with passive bandpass filters. This was inspired by semi-empirical techniques commonly used with magnetic devices [42], [43]. Nevertheless, the results in [32] were only based on heuristic simulations without an experimental verification nor a theoretical analysis that can help maximize the BW. In this paper, we develop a rigorous theory for broadband cyclic-symmetric magnetless circulators that not only provides a systematic design procedure to optimize the performance, but it also allows to derive bounds on the maximum possible BW. Furthermore, an experimental validation is presented, achieving a 20 dB IX BW of 13.9% which is six times larger than our previously reported results [28]-[32].

This paper is organized as follows. In Section II, we develop the theory of broadband STM circulators. Without any loss of generality, we focus on the differential current-mode topology [29] keeping in mind that the analysis can be extended in a straightforward manner to any cyclic-symmetric, passive, and IM-free circulator. In the same section, we find a bound on the maximum BW that can be achieved and study the impact of changing the STM biasing parameters. In Section III, we present simulated and measured results for a PCB prototype at 1 GHz. Finally, we draw our conclusions in Section V.

## II. THEORY

### A. Network Analysis of the Non-Reciprocal Junction

Fig. 1(a) shows the proposed broadband magnetless circulator, which consists of a narrowband STM junction and three identical bandpass filters, one at each port of the junction. We will assume in this paper that the narrowband junction is based on the differential current-mode topology [29] shown in Fig. 1(b), wherein the variable capacitance $C_n$ is given by

Fig. 2. Magnetless circulator terminated with characteristic admittance $Y_c$, resulting in infinite IX at all frequencies.

$$C_n = \begin{cases} C_0 + \Delta C \cos(\omega_m t + (n-1)2\pi/3), & n = 1,2,3 \\ C_0 - \Delta C \cos(\omega_m t + (n-1)2\pi/3), & n = 4,5,6 \end{cases} \quad (1)$$

where $C_0$ is the static capacitance and $\Delta C$ and $\omega_m = 2\pi f_m$ are the modulation amplitude and frequency, respectively. Also, $L_0$ and $R_0 = \omega_0 L_0 / Q_0$ are the total inductance and resistive loss of each tank, respectively, $Q_0$ is the unloaded quality factor, and $\omega_0 = 2\pi f_0 = 1/\sqrt{L_0 C_0}$ is the circulator center frequency. We also assume that the bandpass filters are built using LC tanks in a ladder configuration, as shown in Fig. 1(c).

Using Thevenin equivalence principle, the filters and the 50 Ohm ports in Fig. 1(a) can be replaced by the voltage source $V_s$ and the complex admittance

$$Y_c = 1/Z_c = G + jB, \quad (2)$$

as shown in Fig. 2 where $G$ and $B$ are the real and imaginary parts of $Y_c$, respectively, and they are frequency dispersive quantities. Furthermore, the STM junction can be characterized independently of the loads connected at its ports using the $Y$-parameters, which can be calculated as

$$\bar{\bar{Y}} = Y_0 \left(\bar{\bar{U}} + \bar{\bar{S}}\right)^{-1} \left(\bar{\bar{U}} - \bar{\bar{S}}\right) = \begin{bmatrix} Y_{11} & Y_{31} & Y_{21} \\ Y_{21} & Y_{11} & Y_{31} \\ Y_{31} & Y_{21} & Y_{11} \end{bmatrix}, \quad (3)$$

where $Y_0 = 1/Z_0 = 1/50$ ℧, $\bar{\bar{U}}$ is the unitary matrix, and $\bar{\bar{S}}$ is the $S$-matrix, which was derived in [29] as a function of the circuit elements $L_0$, $C_0$, $Q_0$ and the modulation parameters $f_m$ and $\Delta C$. For convenience, closed-form analytical expressions of the matrix coefficients in (3) are provided in Appendix A.

Notice that both $S$- and $Y$-matrices are cyclic-symmetric thanks to the three-fold symmetry of STM circulators. It is also assumed that these matrices satisfy the passivity condition (see Appendix B), i.e., there is no power exchange between the modulation and the RF signals which is true for modulation frequencies sufficiently below the parametric oscillation condition $f_m = 2f_{rf}$. Moreover, (3) relates the fundamental harmonics at the different ports, yet finite intermodulation (IM) products may also be present. Nevertheless, these products are sufficiently small in differential architectures, and they are entirely cancelled in the ideally symmetric differential configuration [29], therefore they can be neglected. It is worth highlighting that the cyclic-symmetry, passivity, and IM-free features of the junction are, in fact, necessary and sufficient conditions to permit a linear time-invariant (LTI) network analysis, which is the basis behind the proposed BW extension technique. Therefore, this technique can be extended to any non-reciprocal circuit that satisfies these three conditions.

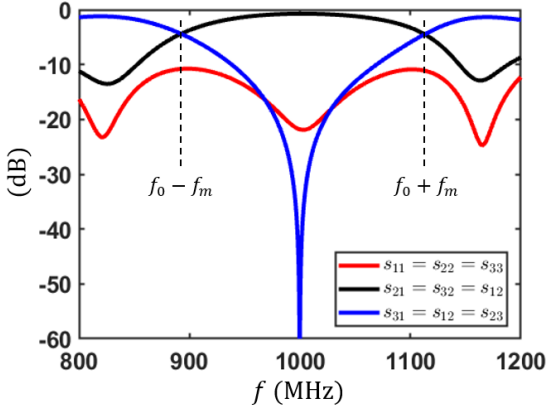

Fig. 3. Theoretical *S*-parameters of the STM junction based on the values provided in Table I.

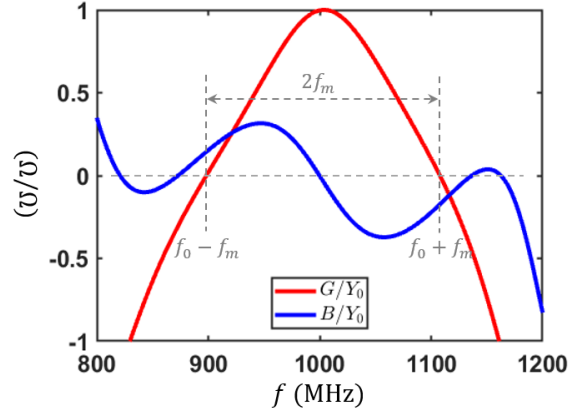

Fig. 4. Normalized real (*G*) and imaginary (*B*) parts of the characteristic admittance $Y_c$ versus frequency.

TABLE I
THEORETICAL DESIGN PARAMETERS OF THE STM JUNCTION.

| Element | Value |
|---|---|
| $L_0$ | 25 nH |
| $C_0$ | 1.2 pF |
| $\Delta C/C_0$ | 54.4% |
| $f_m/f_0$ | 11% |

Using Kirchhoff's laws, the port voltages $\bar{V} = \{V_1, V_2, V_3\}$ in Fig. 2 can be related to the applied voltage sources $\bar{V}_s = \{V_{s1}, V_{s2}, V_{s3}\}$ using

$$\bar{V} = \left(\bar{\bar{U}} + Z_c \bar{\bar{Y}}\right)^{-1} \bar{V}_s. \quad (4)$$

For simplicity, we assume that only port 1 is excited, i.e., $\bar{V}_s = \{1,0,0\}$. A general excitation at all ports can be constructed from a superposition of individual excitations at each port, whose solutions can be incurred from the solution of $\bar{V}_s = \{1,0,0\}$ by using the cyclic symmetry of the device. Substituting (3) into (4) yields

$$\frac{V_3}{V_s} = Y_c \frac{Y_{21}^2 - (Y_c + Y_{11})Y_{31}}{\left\|\bar{\bar{Y}}\right\| + Y_c^3 + 3Y_{11}Y_c^2 + 3(Y_{11}^2 - Y_{21}Y_{31})Y_c}, \quad (5)$$

where

$$\left\|\bar{\bar{Y}}\right\| = Y_{11}^3 + Y_{21}^3 + Y_{31}^3 - 3Y_{11}Y_{21}Y_{31}. \quad (6)$$

Similar relations can also be found for voltages at the other ports, but they are not necessary for our analysis. For proper operation of the circulator, the IX of the combined network is required to be larger than 20 dB at least, therefore we can assume $V_3 \approx 0$, then (5) yields

$$Y_c = \frac{Y_{21}^2}{Y_{31}} - Y_{11}. \quad (7)$$

Furthermore, IL must be maintained below 3 dB to strengthen the argument of full-duplex radios. To simplify the analysis, we assume that the junction's contribution to the total IL is negligible and will account for it in simulations.

Obviously, (7) can only be satisfied over a finite frequency range, which limits the circulator's BW. To explain this further, consider a non-reciprocal junction designed using the values of circuit elements and modulation parameters provided in Table I, and assuming $Q_0 = 50$. These values were chosen such that the junction can provide large isolation (>60 dB) at $f_0 = 1$ GHz if connected to 50 Ohm ports, as shown in Fig. 3 [29],[32]. Notice that we used lowercase letters for the *S*-parameters of the junction itself, whereas uppercase letters will be used for the *S*-parameters of the combined network. The instantaneous BW of the 50 Ohm terminated junction is 4% and insertion loss and return loss at $f_0$ are 0.74 dB and 22 dB, respectively. The junction *Y*-matrix can now be calculated at all frequencies using the analytical expressions in Appendix A. Consequently, the characteristic admittance $Y_c$ can be calculated using (7), where the real (*G*) and imaginary (*B*) parts are shown versus frequency in Fig. 4. At the center frequency $f_0$, *G* is equal to $Y_0$ while *B* is equal to zero, i.e., the junction is perfectly matched to an admittance $Y_0$, which explains the large isolation and low insertion loss at $f_0$ in Fig. 3. However, as the frequency deviates from $f_0$, *B* becomes non-zero and *G* decreases until it becomes negative at $f_0 \pm f_m$, since the direction of circulation is reversed, i.e., transmission follows $1 \rightarrow 3 \rightarrow 2$ rather than $1 \to 2 \to 3$, as shown in Fig. 3. Since $Y_c$ must be causal and passive, the circulator BW is limited to $2f_m$, i.e.,

$$\text{BW} < 2f_m. \tag{8}$$

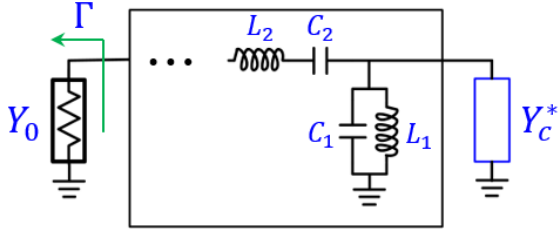

Fig. 5. Synthesis of the bandpass filters in Fig. 1 by matching the RF ports to the complex conjugate of the junction's characteristic admittance $Y_c$.

TABLE II
THEORETICAL DESIGN PARAMETERS OF THE FILTER ELEMENTS.

| Element | Value |
|---|---|
| $L_1$ | 1.76 nH |
| $C_1$ | 15.6 nH |
| $L_2$ | 10.4 pF |
| $C_2$ | 3.90 pF |

Equation (8) shows that increasing the modulation frequency enables a larger BW. However, a smaller $f_m$ may still be desirable to reduce the dynamic power consumption and to permit integration using high-voltage thick-oxide CMOS technologies that can handle high power. In this paper, we choose $f_m/f_0 = 11\%$ for a fair comparison with the narrowband results (without filters) in [32].

B. *Synthesis of the Bandpass Filters and Bode-Fano Limits*

The input admittance $Y_{in}$ seen at port 1 of the non-reciprocal junction can be calculated as

$$Y_{in} = \frac{I_1}{V_1} = Y_{11} - \frac{Y_{31}^2}{Y_{21}}. \tag{9}$$

It is important to stress that (9) is calculated under the assumption that ports 2 and 3 of the junction are terminated with $Y_c$, which is a necessary condition to maximize the circulator BW, as discussed in Section II.A. In the limiting case of a lossless junction, it can be shown that $Y_{in} = Y_c^*$ (see Appendix B), i.e., broadening the BW of the non-reciprocal junction is essentially the same as conjugate matching a three-port network. Following this conclusion, the required bandpass filters in Fig. 1(a) can be designed using the one-port equivalent problem shown in Fig. 5, wherein the goal is to deliver maximum power to a load admittance $Y_c^*$. In order to achieve this goal, reflections at the 50 Ohm ports must be minimized, which implies that the Bode-Fano limits [44] of conventional reciprocal networks also apply to cyclic-symmetric magnetless

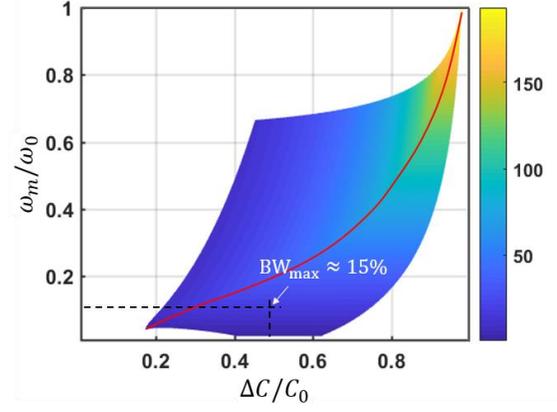

Fig. 6. Maximum BW of the broadband current-mode STM circulator shown in Fig. 1 versus the modulation parameters ($f_m/f_0$ and $\Delta C/C_0$).

circulators. This can be demonstrated analytically by developing a first-order circuit model for the admittance $Y_c^*$, through inspection of Fig. 4, which consists of a series *RLC* tank, the values of its elements are calculated as

$$R_c = |Z_c|_{\omega=\omega_c} \tag{10}$$

$$L_c = 0.5 \left| \frac{dZ_c}{d\omega} \right|_{\omega=\omega_c} \tag{11}$$

$$C_c = \frac{1}{\omega_c^2 L_c}. \tag{12}$$

where $\omega_c = 2\pi f_c$ is the resonant frequency of $Z_c$ which is approximately equal to the circulator's center frequency $\omega_0$. The Bode-Fano criterion then requires

$$\int_0^\infty \ln\left(\frac{1}{|\Gamma|}\right) d\omega \leq \frac{\pi R_c}{L_c}, \tag{13}$$

where $\Gamma$ is the input reflection coefficient depicted in Fig. 5. Notice that the circulator's RL is related to $\Gamma$ as follows $\text{RL} = -20\log_{10}(|\Gamma|)$. Also, for IL less than $\alpha$ and IX larger than $\beta$, power conservation requires

$$\text{RL} > \rho = -20\log_{10}\left(1 - 10^{-\beta/20} - 10^{-\alpha/20}\right), \tag{14}$$

where power dissipation was neglected for simplicity. For $\alpha = 3$ dB and $\beta = 20$ dB, a simple substitution in (14) yields $\rho = 14.33$ dB. Assuming brickwall matching with a constant reflection coefficient $\rho$ inside the BW and unity elsewhere, then (13) yields

$$\text{BW} \leq \frac{\pi}{\ln(1/\rho)} \frac{f_c}{Q_c}, \tag{15}$$

where

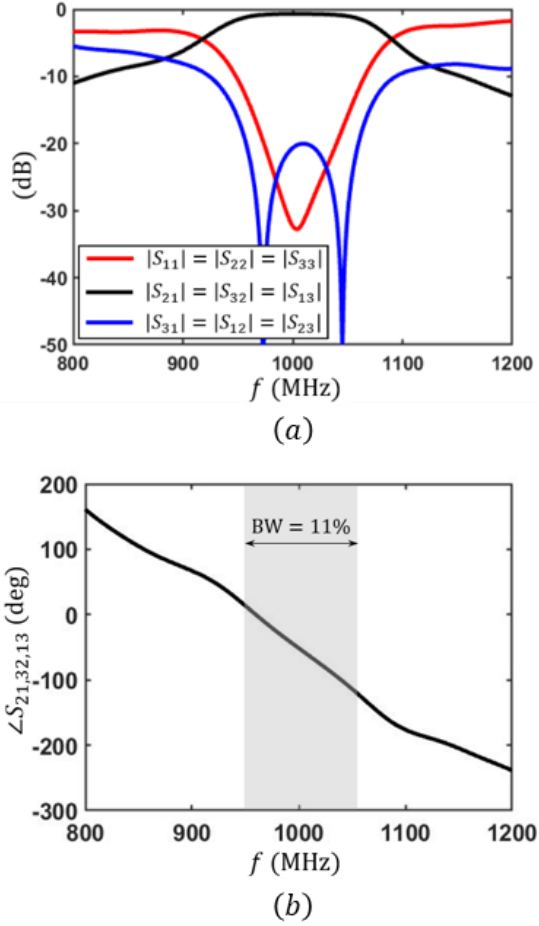

Fig. 7. *S*-parameters of the broadband circulator based on the design parameters provided in Table I and Table II: (a) Magnitude. (b) Transmission phase.

TABLE III
LIST OF THE COMPONENTS VALUES AND PARAMETERS USED IN THE DESIGN OF THE BROADBAND CIRCULATOR.

| | Element | Value |
|---|---|---|
| Junction | $D$ | ~ 1 pF @ VDC=8V |
| | $L_0$ | 24 nH |
| | $V_m$ | 2.2 Vrms |
| | $f_m$ | 110 MHz |
| Filters | $L_1$ | 2.4 nH |
| | $C_1$ | 7.8 pF |
| | $L_2$ | 11 nH |
| | $C_2$ | 2.5 pF |
| DC and modulation network | $R_b$ | 100 KOhm |
| | $C_b$ | 1000 pF |
| | $L_d$ | 68 nH |
| | $L_m$ | 220 nH |
| | $C_m$ | 5 pF |

$$Q_c = \frac{1}{\omega_c R_c C_c} = \frac{\omega_c L_c}{R_c}. \quad (16)$$

From (8) and (15), we get

$$\text{BW} < \min\left\{\frac{\pi}{\ln(1/\rho)}\frac{f_c}{Q_c}, 2f_m\right\}. \quad (17)$$

Equation (17) provides a global bound on the BW of any passive pseudo-LTI cyclic-symmetric magnetless circulator. For the special case under consideration of a current-mode STM circulator, Fig. 6 shows the impact of changing the modulation parameters on this bound. The colorful portion of the figure depicts the possible combinations of $f_m/f_0$ and $\Delta C/C_0$ that could achieve an overall IX larger than 20 dB and a total IL less than 3 dB. Other combinations, however, as depicted by the white region cannot meet these specs, hence the BW definition in this case becomes invalid. Within the valid range, Fig. 6 also shows that there is an optimal one-to-one mapping between the modulation parameters $f_m/f_0$ and $\Delta C/C_0$ that maximizes the circulator's BW (smallest possible modulation parameters for a given BW), as indicated by the red line. For the parameters listed in Table I, i.e. $f_m/f_0 = 11\%$ and $\Delta C/C_0 = 54.4\%$, the maximum possible BW is about 15%. Obviously, a smaller $\Delta C/C_0$ would have resulted in a larger BW for the same $f_m/f_0 = 11\%$, which could not have been predicted by heuristic simulations. This can be corrected in future designs, while here we will continue to assume the same parameters in Table I and design the matching network to approach the 15% bound. By observing that the load admittance $Y_c^*$ is resonant at the circulator's center frequency (see Fig. 4), then the required matching networks can be reduced to simple *LC* bandpass filters as shown in Fig. 5. Obviously, these filters increase the circulator's overall insertion loss; therefore, their order should be kept small (<3 in practice).

Assuming a second-order filter and initially ignoring losses to simplify the design, the input impedance $Z_{in}$ seen by the 50 Ohm ports in Fig. 5 can be calculated as follows

$$Z_{in} = Z_2 + \frac{1}{Y_1 + Y_c^*}, \quad (18)$$

where

$$Y_1 = j\left(\omega C_1 - \frac{1}{\omega L_1}\right), \quad (19)$$

$$Z_2 = j\left(\omega L_2 - \frac{1}{\omega C_2}\right), \quad (20)$$

and $L_k$ and $C_k$ are the inductance and capacitance of the ladder's *k*-th branch, respectively, where $k = 1$ at the side of $Y_c^*$. Assuming a Chebyshev response, since it gives larger BW than Butterworth's while Elliptic and other responses are not worth the additional complications as noted in [44], then the filter elements can be calculated by enforcing the impedance conditions

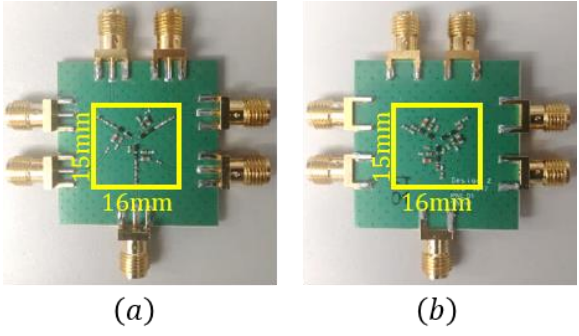

Fig. 8. Photographs of the fabricated broadband STM circulator prototype: (a) Top side. (b) Bottom side.

$$Z_{in}(f_c \pm \delta f/2) = Z_0, \qquad (21)$$

where $f_c = f|_{B=0} \approx f_0$ is the resonant frequency of $Y_c^*$ and $\delta f$ is chosen numerically to maximize the circulator's BW within the global bound given by (17). Thereafter, the S-parameters of the broadband circulator can be calculated as shown in Fig. 7 using signal-flow graph analysis as explained in Appendix C. Clearly, the IX follows a second-order Chebyshev response as expected with two notches at 972 MHz and 1045 MHz and in-band ripple less than 20 dB. Furthermore, the achieved BW is 11% which is about 2.5 times larger than the narrowband junction's 4% and at least 4.5 times larger than the results reported in [28]-[32]. This value can be improved further, approaching the limit of 15% in Fig. 6, by relaxing the conditions in (21) and not necessarily requiring IX to be infinite at $f_c \pm \delta f/2$, as long as it is still larger than 20 dB. This will be shown through simulations in the next section while taking into account the neglected dissipation of the filter elements and all parasitics.

## III. RESULTS AND DISCUSSION

In Section II, dissipation of the filters was neglected. In practice, this nuisance not only degrades IL, but it also distorts the filter characteristics, hence the synthesized IX dispersion is perturbed. In order to account for these effects, circuit/EM co-simulations were performed in Keysight ADS while tweaking the design parameters to compensate for all parasitic effects, thus resulting in the final values summarized in Table III. The design was then fabricated on a PCB using discrete off-the-shelf components as shown in Fig. 8 and the measurements were taken using the same experimental setup in [28] and [29].

Fig. 9 shows the simulated and measured S-parameters of the narrowband junction (without filters). The simulated BW in this case is 4.1% (41 MHz), within which IX from the TX to the RX (TX-RX IX) is larger than 20 dB, IL from the TX to the ANT (TX-ANT IL) is identical to IL from the ANT to the RX (ANT-RX IL) and both vary from 1.5 dB to 1.75 dB. RL at the ANT (ANT RL) is also identical to RL at the TX (TX RL) and both are less than 16 dB. Similarly, the measured BW is about 4% (40 MHz) wherein TX-RX IX is larger than 20 dB, TX-ANT IL is less than 2.2 dB, ANT-RX IL is less 2.3 dB, while ANT RL and TX RL are less than 14.6 dB and 14 dB, respectively. Simulated and measured results are in fair agreement although the measured IL is about 0.5~0.6 dB larger than the simulated value due to inaccuracies in the commercially available spice models of the used components. The measured S-parameters also exhibit finite asymmetry, which is more prominent in the RL, due to finite tolerance of the components and layout effects.

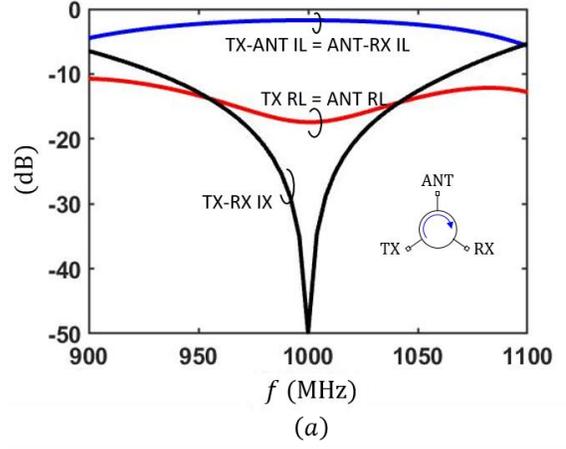

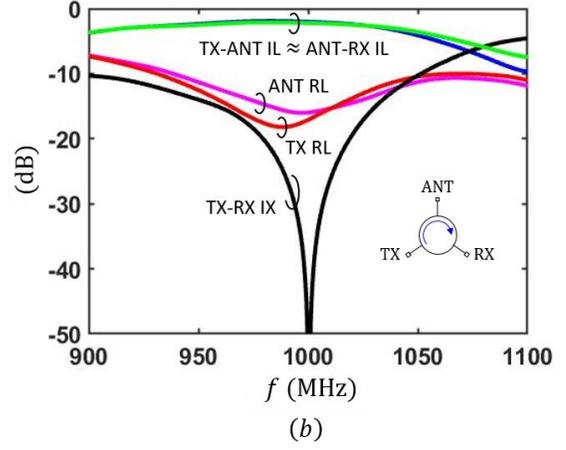

Fig. 9. S-parameters of the narrowband junction (without filters): (a) Simulated. (b) Measured.

Fig. 10 shows the simulated and the measured S-parameters of the broadband circulator, i.e. after adding the filters. In this case, the simulated and the measured BW are 12% (120 MHz) and 13.9% (139 MHz), respectively, i.e. the circulator's BW increased by a factor of three, approximately, compared to the narrowband junction. The simulated in-band IL for both the TX-ANT and the ANT-RX varies from 2.5 dB at the center frequency to 4 dB at the band edges while RL at all ports varies from 33.4 dB to 11 dB. Compared to the narrowband junction, IL of the broadband circuit increases by about 1 dB due to the added filters (each filter contributes 0.5 dB) while degradation at the edges is larger simply because the 20 dB IX BW has increased. Should IL be limited to 3 dB as required in practice, then the BW effectively reduces to 7.4% which is still 1.85 times larger than the junction's BW. This problem can be mitigated by using distributed elements, at the expense of increasing the form factor. Another option is to use MEMS filters which have high quality factors and at the same time a small size [33]-[38]. On the other hand, Fig. 10(b) shows that the measured in-band TX-ANT IL varies from 4.2 dB to 5.8 dB, ANT-RX IL varies from 4.25 dB to 5.5 dB, TX-RX IX is larger than 20 dB, while TX RL and ANT RL are better than 12.6 dB and 11.7 dB, respectively. Similar to the narrowband junction, the high IL and finite asymmetry are mainly due to model inaccuracies, particularly for the varactors. More importantly, the finite tolerance of the lumped elements in the filters distorts their characteristics and increases reflections, thus further degrading the overall IL. These problems can be overcome by further optimizing the layout and incorporating more accurate models of the components, particularly the varactors. Finally, Table IV summarizes the achieved results in comparison to previous works on STM circulators. It is also worth mentioning that the impact of the presented BW extension method on the power handling, linearity, and harmonic response of STM circulators is negligible. Therefore, these metrics show similar performance to the narrowband junction [32] and can be improved by using similar approaches to those discussed in [28], [29]. The noise figure, on the other hand, is degraded in proportion to the additional losses incurred by the filters. Therefore, improving the overall IL as explained earlier would improve the broadband circulator's noise figure as well.

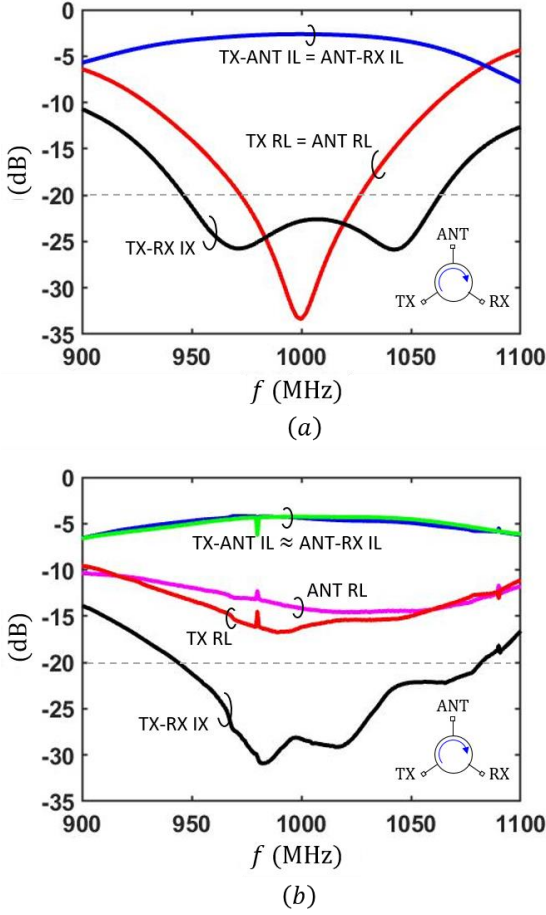

Fig. 10. *S*-parameters of the broadband circulator (with filters): (a) Simulated. (b) Measured.

TABLE IV
Summary of the Measured *S*-parameters in Comparison to Previous Works on STM Circulators.

| Metric | This work | [32] | [29] | [28] |
|---|---|---|---|---|
| RF Center Freq. (MHz) | 1000 | 1000 | 1000 | 1000 |
| Mod/RF Freq. Ratio (%) | 11 | 11 | 10 | 19 |
| 20 dB IX BW (%) | 13.9 | 4 | 2.3 | 2.4 |
| TX-ANT IL (dB) | 4.2~5.8 | 1.9~2.2 | 1.78~2 | 3.3~3.56 |
| ANT-RX IL (dB) | 4.25~5.5 | 2.1~2.3 | 1.78~2 | 3.3~3.56 |
| TX RL (dB) | 12.6~16.7 | 14.6~18 | 20~23 | 9~11.3 |
| ANT RL (dB) | 11.7~14.6 | 14~16 | 20~23 | 9~11.3 |

## IV. CONCLUSION

We presented broadband magnet-less circulators based on combining STM narrowband junctions with three identical bandpass filters, one at each port of the junction. We developed a rigorous theory for such circuits, which enabled a systematic design of the constituent elements and, more importantly, allowed to derive a global bound on the maximum possible BW. Guided by the developed theory, a PCB prototype was designed, resulting in a measured 20 dB IX BW of 13.9%, which is 5.8 times larger than all previous results on STM circulators [28], [29]. While the junction's electrical tunability is sacrificed (recall that STM circulators can be tuned over different channels by controlling the DC bias [28]), the improvement in BW is actually drastic that it exceeds the whole tunability range of the narrowband junction, thus making it unnecessary to maintain this characteristic. Also, the in-band IL can be easily improved to less than 3 dB by further optimizing the layout and by using low-loss filters based on distributed elements or compact high-*Q* MEMS resonators.

## APPENDIXES

### A. *Y-parameters of the Non-Reciprocal Junction*

Here, we provide analytical expressions for the *Y*-parameters of the STM differential current-mode junction as function of circuit elements and modulation parameters, viz.

$$Y_{11}(\omega) = 2\left[I_+(\omega) + I_-(\omega)\right] \quad (22)$$

$$Y_{21}(\omega) = 2\left[e^{+j2\pi/3}I_+(\omega) + e^{-j2\pi/3}I_-(\omega)\right] \quad (23)$$

$$Y_{31}(\omega) = 2\left[e^{-j2\pi/3}I_+(\omega) + e^{+j2\pi/3}I_-(\omega)\right], \quad (24)$$

where

$$I_\pm(\omega) = \frac{-j\omega}{3L_0}\left[\omega^2 - j\frac{\omega_0}{Q_0}\omega - \sigma\omega_0^2 - \frac{\omega_0^4\gamma^2/4}{(\omega\pm\omega_m)^2 - \sigma\omega_0^2 - j(\omega_0/Q_0)(\omega\pm\omega_m)}\right]^{-1} \quad (25)$$

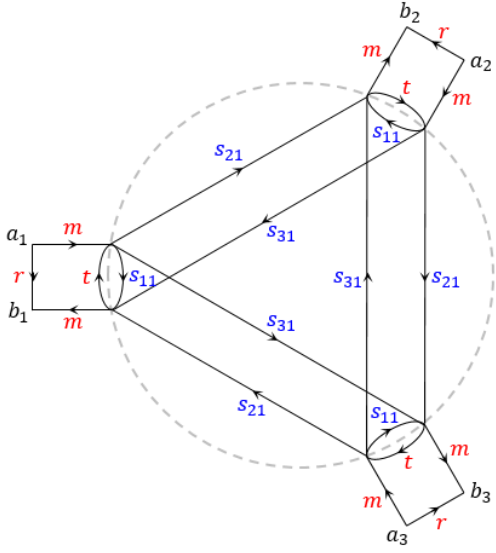

Fig. 11. Signal-flow graph of the broadband circulator shown in Fig. 1(a).

$$\sigma = \frac{1}{\sqrt{1-(\Delta C/C_0)^2}} \quad (26)$$

$$\gamma = \frac{2C_0}{\Delta C}\left(1 - \frac{1}{\sqrt{1-(\Delta C/C_0)^2}}\right). \quad (27)$$

Equations (22)-(27) can be derived following a similar analysis to [29].

### B. Input Admittance of the Non-Reciprocal Junction

In this Appendix section, we show that the input admittance $Y_{in}$ of the non-reciprocal junction is equal to $Y_c^*$, when terminated with $Y_c$ at all ports and losses are neglected. Passivity of LTI $N$-port networks requires

$$\sum_{n=1}^{N} P_n \geq 0, \quad (28)$$

where $n$ is the port index and $P_n$ is the input power at the $n$-th port. Equation (28) can be rewritten in a matrix form as follows

$$\mathrm{Re}\{\bar{V}^\dagger \bar{I}\} \geq 0, \quad (29)$$

where $\bar{V} = \{V_1, V_2, V_3\}$ and $\bar{I} = \{I_1, I_2, I_3\}$ are the vectors of the voltages and currents at the ports, respectively, and $\dagger$ is the conjugate transpose operator. Substituting for $\bar{I} = \bar{\bar{Y}}\bar{V}$ into (29) and assuming a lossless and cyclic-symmetric junction, then (29) simplifies to

$$Y_{11} + Y_{11}^* = 0 \quad (30)$$
$$Y_{21} + Y_{31}^* = 0. \quad (31)$$

Substituting (30) and (31) into (9) results in

$$Y_{in} = -Y_{11}^* + \frac{(Y_{21}^*)^2}{Y_{31}^*} = Y_c^*, \quad (32)$$

where $Y_c$ is given by (7).

### C. Signal-Flow Graph Analysis of the Broadband Circulator

In this Appendix section, we derive analytical expressions for the $S$-parameters of the broadband circulator (junction + filters) based on signal-flow graph analysis. First, the $S$-parameters of the junction can be written in the form

$$\bar{\bar{S}} = \begin{bmatrix} s_{11} & s_{31} & s_{21} \\ s_{21} & s_{11} & s_{31} \\ s_{31} & s_{21} & s_{11} \end{bmatrix}. \quad (33)$$

Also, the $S$-parameters of the filters can be calculated as follows

$$S = \begin{bmatrix} r & m \\ m & t \end{bmatrix}, \quad (34)$$

where

$$r = 1 - \frac{2Z_0(Y_0 + Y_1)}{1 + (Y_0 + Y_1)(Z_0 + Z_2)} \quad (35)$$

$$t = 1 - 2Z_0 \frac{1 + Y_1(Z_0 + Z_2)}{2Z_0 + Z_2 + Z_0 Y_1(Z_0 + Z_2)} \quad (36)$$

$$m = \frac{2}{1 + (Y_0 + Y_1)(Z_0 + Z_2)}, \quad (37)$$

where $Y_1$ and $Z_2$ are given by (19) and (20), respectively.

Fig. 11 shows the signal-flow graph of the combined network where $a_i$ and $b_i$ are the incident and reflected wave amplitudes at the $i$-th port, respectively. Using Mason's gain formula [45], the $S$-parameters of the combined network can be calculated as follows

$$S_{11} = S_{22} = S_{33} = r + \frac{\Delta_1}{\Delta}m^2 s_{11} + \frac{2}{\Delta}m^2 s_{21}s_{31}t(1-s_{11}t) + \frac{1}{\Delta}m^2 s_{21}^3 t^2 \quad (38)$$

$$S_{21} = S_{32} = S_{13} = \frac{1}{\Delta}m^2\left[s_{21}(1-s_{11}t) + s_{31}^2 t\right] \quad (39)$$

$$S_{31} = S_{12} = S_{23} = \frac{1}{\Delta}m^2\left[s_{31}(1-s_{11}t) + s_{21}^2 t\right], \quad (40)$$

where

$$\Delta_1 = 1 + s_{11}^2 t^2 - (2s_{11}t + s_{21}s_{31}t^2) \quad (41)$$

$$\Delta = 1 - s_{11}^3 t^3 + (3s_{11}^2 t^2 + 3s_{11}s_{21}s_{31}t^3) - (3s_{11}t + 3s_{21}s_{31}t^2 + s_{21}^3 t^3). \quad (42)$$